# Principles for Designing Computer Music Controllers

**Perry Cook**
Department of Computer Science (also Department of Music)
35 Olden St.   Princeton, NJ 08544 USA
+1 609 258 4951      prc@cs.princeton.edu

**ABSTRACT**
This paper will present observations on the design, artistic, and human factors of creating digital music controllers. Specific projects will be presented, and a set of design principles will be supported from those examples.

**INTRODUCTION**
Musical performance with entirely new types of computer instruments is now commonplace, as a result of the availability of inexpensive computing hardware, of new sensors for measuring physical parameters such as force and position, and of new software for real-time sound synthesis and manipulation. Musical interfaces that we construct are influenced greatly by the type of music we like, the music we set out to make, the instruments we already know how to play, and the artists we choose to work with, as well as the available sensors, computers, networks, etc. But the music we create and enable with our new instruments can be even more greatly influenced by our initial design decisions and techniques.

Through designing and constructing controllers over the last 15 years, the author has developed some principles and a (loose) philosophy. These are not assumed to be universal, but are rather a set of opinions formed as part of the process of making many musical interfaces. They relate to practical issues for the modern instrument craftsperson/hacker. Some relate to human factors, others are technical. This paper will endeavor to bring those principles to light, through a set of examples of specific controllers and related art projects. To set the tone for the rest of the paper, the principles will be listed here, and will be highlighted in bold when they are reinforced by the examples in the text.

**Some Human/Artistic Principles**

1) **Programmability is a curse**
2) **Smart instruments are often not smart**
3) **Copying an instrument is dumb, leveraging expert technique is smart**
4) **Some players have spare bandwidth, some do not**
5) **Make a piece, not an instrument or controller**
6) **Instant music, subtlety later**

**Some Technological Principles**

7) **MIDI = Miracle, Industry Designed, (In)adequate**
8) **Batteries, Die (a command, not an observation)**
9) **Wires are not that bad (compared to wireless)**

**Some Other Principles**

10) **New algorithms suggest new controllers**
11) **New controllers suggest new algorithms**
12) **Existing instruments suggest new controllers**
13) **Everyday objects suggest amusing controllers**

**Winds:  Cook/Morrill Trumpet 1986-89      HIRN 1991**
Constructed with Dexter Morrill of Colgate University, as part of an NEA grant to create an interface for trumpeter Wynton Marsalis, the Cook/Morrill trumpet controller project led to a number of new interface devices, software systems [1][2], and musical works [3]. Sensors on the valves, mouthpiece, and bell enabled fast and accurate pitch detection, and extended computer control for the trumpet player. Trumpet players lie squarely in the "**some players have spare bandwidth**" category, so attaching a few extra switches and sliders around the valves proved very successful. Figure 1 shows the interface window.

Initially it was thought that a musically interesting scheme would be to allow the brass player to use the switches to enter played notes into loops, and later trigger those loops. This proved a miserable failure, because of the mental concentration needed to keep track of which loop was where, what the loop contents were, syncing the recording, triggering, etc. Eventually a set of simple, nearly stateless interactions were devised. The switches were used to trigger pre-composed motifs, navigate forward and backward through sections, and capture pitch information from the horn, which was then used to seed fairly autonomous compositional algorithms.

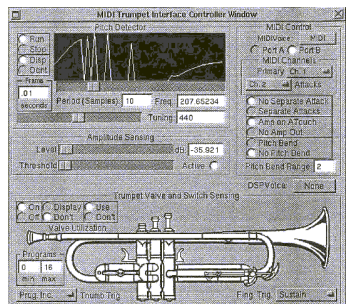

Figure 1   Interface panel for the Cook/Morrill Trumpet

Another project, the HIRN wind controller, sensed rotation and translation in both hands, arm orientation, independent control with each finger, breath pressure, and even muscle tension in the lips [4]. Mappings from these controls to the parameters of the WhirlWind meta-wind-instrument physical model allowed exploration of new "spaces" of acoustical processes, and the HIRN also was investigated as a controller for FM and other synthesis techniques. Negative lessons from the HIRN project indicated that huge control bandwidth is not necessarily a good thing, and that attempting to build a "super instrument" **with no specific musical composition to directly drive the project** (**principle 5**) yields interesting research questions, but with no real product or future direction. One positive lesson from the project is that the **co-design of synthesis/processing algorithms with controllers** can benefit both.

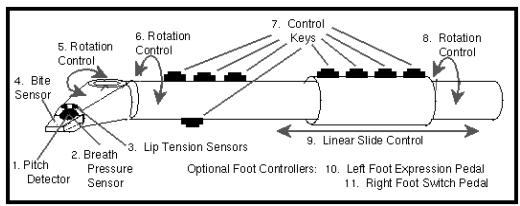

Figure 2: The HIRN Meta-Wind Controller

**Voice: SPASM 1988-94**
Research on physical modeling of the voice resulted in the construction of the SPASM/Singer voice synthesizer [5] [6] (see Figures 3 and 4). The SPASM system was capable of real time synthesis, but had well over 40 continuously controlled parameters. Work to improve the graphical interfaces and add control via MIDI fader boxes [7] proved that the voice is a truly difficult "instrument" to control (**principles 3 & 4**). Recent work in real-time vocal model control will be discussed in the SqueezeVox Section.

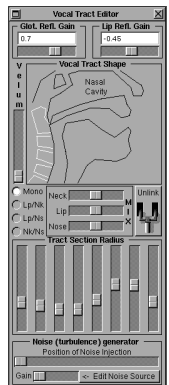 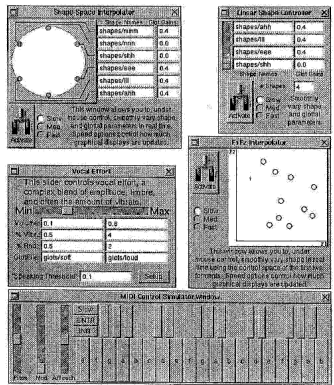

Figure 3 SPASM    Figure 4 Few-to-Many Mappings

**PhISEM Shaker Percussion: 1996-1999**
The PhISEM (Physically Inspired Stochastic Event Modeling) project [8][9] provided support for the "**new algorithms lead to new controllers lead to new algorithms …**" principles. This work on the synthesis of particle-type percussion and real-world sounds led to a set of new instruments, not only for control of shaker/scraper sounds and sound effects, but also for algorithmic interactive music. For example, the Frog Maraca (Figure 5) sends MIDI commands to control a simple algorithmic fusion jazz combo of bass, piano, and drums. The success with both adults and children [10] of the Frog Maraca came from its simple interface (just shake it), the fun of making fairly complicated music with such a simple and whimsical looking device, and the fact that it only performed one function (and always performed that function when turned on). A related shaker percussion instrument controller was a Tambourine that could also compose algorithmic modal marimba solos when shaken.

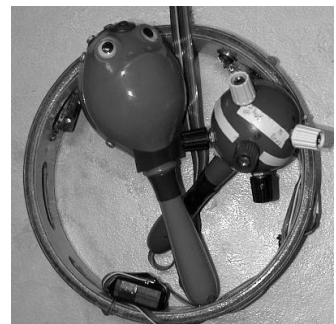

Figure 5 PhISEM controllers.

Constructing the PhISEM controllers provided rich evidence that "**Programmability is a curse**," and a correlary: "**Smart instruments are often not smart**." What these principles are meant to address is that the programmability of computer-based musical systems often make them too easy to configure, redefine, remap, etc. For programmers and composers, this provides an infinite landscape for experimentation, creativity, writing papers, wasting time, and never actually completing any art projects or compositions. For normal humans, being able to pick up an instrument with a simple obvious interaction and "play it" only makes logical sense. That the instrument is "learning from their play and modifying its behavior" often does not make any sense at all, and can be frustrating, paralyzing, or offensive. PhISEM controllers have a single embedded microcontroller, programmed for one or two functions (selectable by the state of a button on power-up), and they put out standard General **MIDI** signals. Except for the need to replace batteries (**Die Batteries Die!!**), these controllers have a strong possibility of working perfectly as designed in 10 (perhaps 20) years. Those who craft complex systems using custom hardware, multiple computers, and multiple operating systems, can make no such claims.

**Foot, Hand, Kitchen Wear/Ware 1997-2000**
Spurred by the success of the PhISEM controllers, the notion of simple, MIDI based, fixed single function controllers was continued, but based on objects that are not specifically associated with music. Figure 6 shows the TapShoe, constructed at Interval Research as part of Bob Adams' Expressions Project. This shoe used force sensing resistors and accelerometers attached directly to a DSP board running PhISEM shaker algorithms and a small rhythmic loop. The algorithm generated a basic "groove" to which the wearer of the shoe could add accents and dynamics, in addition to their own tapping sounds. The success of the system came from giving the TapShoe wearer that feeling that they were actually performing the music, though the algorithmic loop would play a relatively boring tapping sound even if the shoe sat unworn ("**Instant music, subtlety later**").

The Pico Glove (see Figure 7) was **designed as a single composition**, called "Pico I for Seashells and Interactive Glove" [11]. The idiomatic gesture of moving the hand in and out of the shells was enhanced by a tilt sensor in the glove. This was used to steer fractal note-generation algorithms in real time, to accompany the blown shells.

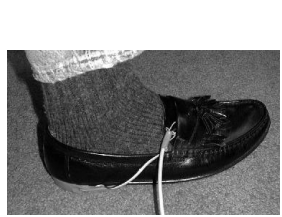 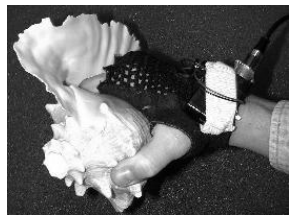

Figure 6  Digital Tapshoe        Figure 7 PicoGlove

The JavaMug (Figure 8) was designed for a transcontinental MIDI jam session held in 1997 between Tokyo and Columbia University [12]. Being one of the author's favorite objects, the coffee mug fits comfortably into the hand, and pressure sensors beneath the fingers, a tilt sensor, a pot and two buttons allow control of an algorithmic techno-latin band. The principle of "**Instant music, subtlety later**" is dominant in this instrument. Simply picking up the JavaMug and squeezing it yields attractive and (fairly) deterministic music, because algorithmic randomness is increased by *decreasing* pressure on the sensors. After playing the instrument for a while, neophytes grow to more expert levels by realizing that the music gets more varied and interesting if they experiment with the relative pressures and tilts. Note that this is also an example of the "**Smart instruments are often not smart**" principle, in that the instrument doesn't change at all, but rather trains the user to use more gentle and subtle manipulations of the sensors. Other kitchen-related interfaces include the "Fillup Glass," which plays minimalist music loops (**via MIDI**) controlled by sensors in a water glass, and "P-Ray's Café: Table 1" which allows the control of a melodic percussion group by movement of common table-top items (sugar, salt shaker, etc) over the surface of a small table. These objects showed that "**Everyday objects suggest amusing controllers.**"

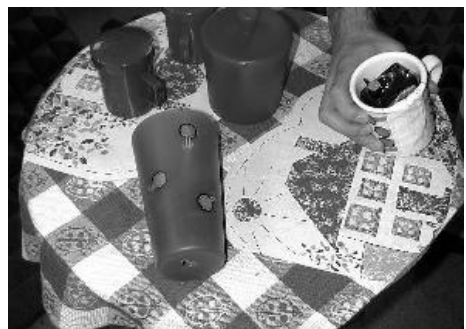

Figure 8   P-Ray's Café, with Fillup Glass and Java Mug

**Violins/Strings:  BoSSA, the Nukelele   1998-99**

Stringed instruments have a rich historical musical tradition. They also have a rich tradition of electronic interfaces, both commercially and experimentally with many electrified, MIDI, and pure digital violins and guitars. Work with Dan Trueman at Princeton University began with a sensor-enhanced violin bow called the "RBow," and the NBody project which worked to study and model the directional radiation properties of stringed instruments[13]. Dan continued and expanded this work, yielding BoSSA (The Bowed Sensor, Speaker Array, Figure 9) [14]. Lessons learned and reinforced by the BoSSA project include "**Existing instruments suggest new controllers**", and "**Copying an instrument is dumb, leveraging expert technique is smart**." Other principles reinforced are "**Some players have spare bandwidth, some do not**," (violin players generally have their hands completely occupied, so a successful interface must exploit interesting remappings of existing gestures), and "**Wires are not that bad (compared to wireless)**" (the BoSSA is played sitting by a player who often plays electric violin, so the increased complexity of wireless was not justified).

The Nukelele (thanks to Michael Brooke for the name) was constructed in Bob Adams' Interval Research Expressions project. While collaborating on other Expressions projects such as "the Stick" and the " Porkophone," the Nukelele was a personal experiment to design, implement, and test a new controller as rapidly as possible. The Nukelele was intended to match the expressiveness of a true stringed instrument, by using audio directly from a sensor to drive a plucked string physical model. Two sandwiched linear force sensing resistors under the right hand served to provide pluck/strike position information, along with the audio excitation for the string model.

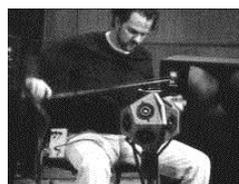 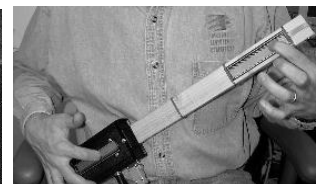

Figure 9   BoSSA              Figure 10, the Nukelele

**The Voice (again): SqueezeVox   2000**

The SqueezeVox project [15] with Colby Leider of Princeton has revisited the difficult issue of devising a suitable controller for models of the human voice. Breathing, pitch, and articulation of vowels and consonants must be controlled in a vocal model, so the accordion was selected as a natural interface (**principle 10**). Pitch via the keyboard, vibrato aftertouch, and a linear strip for fine pitch and vibrato are controlled with the right hand. Breathing is controlled by the bellows, and the left hand controls vowels and consonants via buttons (presets), or continuous controllers such as a touch pad, plungers, or squeeze interface.

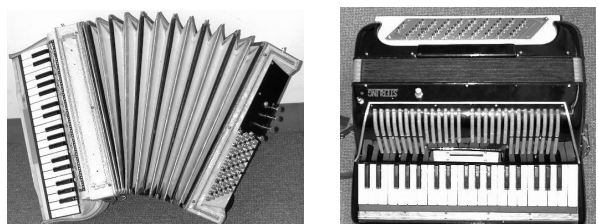

Figure 11 Squeezevox Lisa                and Bart

**FUTURE WORK and CONCLUSIONS**

Work and development continues on the SqueezeVox project, with a self-contained version (Santa's Little Helper, with onboard DSP synthesis), and a small concertina version (Maggie) currently under construction. Work also continues on the kitchen/common objects project, and given the variety of such objects, much rich interface and music design lies ahead.

Musical interface construction proceeds as more art than science, and possibly this is the only way that it can be done. Yet many of the design principles put forth in this paper have held true in multiple projects, and many have been verified in talking with other digital instrument designers. Some of the technological issues might go away, but not completely or not necessarily very quickly. Many of the human/artistic issues are likely to be with us as long as musical instruments have been.

**DEMONSTRATIONS**

During the workshop, the PhISEM controllers, the JavaMug, the TapShoe, the Nukelele, and the SqueezeVox will be demonstrated. Soundfiles, large pictures, and video clips of the instruments discussed in this paper are available at: http://www.cs.princeton.edu/~prc/CHI01.html

**ACKNOWLEDGEMENTS**

Specific thanks to Dexter Morrill, Dan Trueman, Bob Adams, and Colby Leider. General thanks to all those at CCRMA, Princeton, and Interval Research for wonderful collaborations. This work was funded by CCRMA and the CCRMA Industrial Affiliates Program, Interval Research, Intel, and the Arial Foundation.